\documentstyle[prl,aps]{revtex}
\begin{document}
\draft
\twocolumn[\hsize\textwidth\columnwidth\hsize\csname @twocolumnfalse\endcsname

\title{
Phase Separation in the Two-Dimensional Hubbard Model
}
\author{Gang Su$^\ast$
}
\address{
 Institut f\"ur Theoretische Physik,
Universit\"at zu K\"oln\\
 Z\"ulpicher Strasse 77, D-50937 K\"oln, Germany\\
}
\date{Received 30 May 1996, Phys. Rev. {\bf B54}, 8281R(1996)}
\maketitle

\begin{abstract}
By making use of known exact results and symmetry properties for the one-band 
Hubbard
model, we show in a somewhat exact manner that there is 
no phase separation on a square lattice at arbitrary fillings at finite 
temperature for both attractive and repulsive on-site Coulomb interaction. 
This result is consistent with 
the quantum Monto Carlo calculation. 

\end{abstract}

\pacs{PACS numbers: 71.27.+a, 74.80.-g}
]

The considerable experimental evidence shows that $La_2CuO_{4+\delta}$ has
a regime in which phase separation occurs\cite{jorgen}. Such a separation 
appears between a phase very close to $La_2CuO_{4}$ (i.e., an oxygen-poor phase) 
and an oxygen-rich phase that becomes superconducting at $T_c \sim 38K$.
Later on, this phenomenon was actively investigated and confirmed in many other high Tc superconductors\cite{naga}.
People therefore believe that the presence of phase separation is of essential
importance for understanding the physics of the cuprate superconductors\cite{muller}.
Apart from the experimental works, there are also some theoretical 
investigations devoting to this subject, see Refs.\cite{naga,daggo} for excellent reviews.
As most of theoretical studies, being based on the planar t-J and Hubbard models, are numerical
or approximate, the obtained results still remain controversial\cite{daggo,naga}.
Quite recently, Au {\it et al}.\cite{azn}, using the symmetry
properties and known exact results obtained by Lieb\cite{lieb}, Kubo and 
Kishi\cite{kk} for the
Hubbard model, got an exact result on phase separation
in the Hubbard model on bipartite lattices. They asserted that there is no phase separation
at {\it low} dopings at any temperature for the repulsive Hubbard
model. This exact consequence clarifies, to some extent, the existing 
controversy in the Hubbard model and in the t-J model with 
small values of J/t. However, several issues concerning this problem still 
remain to be addressed in the sense that the exact results are sparse. 
For instance, if phase separation exactly occurs in
the cases with moderate or even high dopings or in the attractive (negative-U) Hubbard model
on symmetric bipartite lattices (e.g. on square lattices or cubic lattices), is
still inconclusive. 
Since there have been
numerical results indicating no phase separation for the one-band Hubbard model 
on square lattices at any fillings\cite{moreo,daggo}, it would be quite
interesting to seek more exact evidences supporting this observation.

In this note, making use of known exact results obtained by Ghosh\cite{ghosh}
and symmetry properties for the one-band 
Hubbard model
on a square lattice,  we shall show that there is 
no phase separation at any fillings at finite
temperature for both attractive and repulsive on-site Coulomb interaction. 
This statement, 
being exact, is consistent with the quantum Monto Carlo calculation\cite{moreo,daggo}. 

We start from the one-band Hubbard model in an external field $h$ on a square 
lattice with $M$ sites. The Hamiltonian reads
\begin{eqnarray}
& & H = - t \sum_{<i,j>\sigma}(c_{i\sigma}^{\dagger}c_{j\sigma} + h.c.) + 
U \sum_{i}n_{i\uparrow}n_{i\downarrow}  \nonumber \\ \hspace{0.5cm}
& & - \mu \sum_{i}(n_{i\uparrow} + n_{i\downarrow}) - \frac{h}{2}\sum_{i}(n_{i\uparrow} - n_{i\downarrow}),
\end{eqnarray}
where the notations are standard. The foregoing discussion is independent of 
the sign of $U$, and is only valid for finite temperatures and for the 
finite chemical potential. In the following, 
we shall first investigate 
the phase separation near half-filling, and then discuss the problem away from
half-filling.
The advantage of our method is the fact that we can treat the
problem in a unifying way for both $U>0$ and $U<0$. 

The method which we shall adopt for the case near half-filling, being based on the well-known particle-hole transformation,
is very similar to that exploited in Ref.\cite{azn}, but we do not invoke any
{\it a priori} assumption. 
For the Hamiltonian (1) defined on a square lattice, Ghosh\cite{ghosh} obtained, 
for small $h$, an exact 
upper bound
\begin{eqnarray}
|m(h,T)| \leq \frac{const.}{T^{\frac12}} \frac{1}{|\ln |h||^{\frac12}},
\end{eqnarray}
where $m(h,T) = \frac{1}{M} \sum_i \langle S_i^z \rangle$, the magnetization per
site, and $\langle \cdots \rangle$ is the thermal average over a grand 
canonical ensemble. As $h \rightarrow 0$, $|m(h,T)| \rightarrow 0$ means the
absence of spontaneous magnetic long-range order, implying that the system 
could exhibit paramagnetic behaviors at temperature $T>0$. For small $h$, 
inequality (2) ensures the analyticity (only in the sense that the first 
derivative exists) of $m(h,T)$. 
Therefore, we can conclude
that $m(h,T>0)$ is
continuous and analytic in the neighborhood of $h=0$.

We now apply the unitary particle-hole transformation,
\begin{eqnarray}
c_{i\uparrow} \rightarrow c_{i\uparrow}, ~~~~c_{i\downarrow} \rightarrow 
\epsilon(i) c_{i\downarrow}^{\dagger},
\end{eqnarray}
with $\epsilon(i)=-1$ for $i \in$ the one sublattice and $-1$ for $i \in $ the 
other sublattice, to the Hamiltonian 
(1), and obtain
\begin{eqnarray}
& & H' = - t \sum_{<i,j>\sigma}(c_{i\sigma}^{\dagger}c_{j\sigma} + h.c.) + 
U' \sum_{i}n_{i\uparrow}n_{i\downarrow}  \nonumber \\ \hspace{0.5cm}
& & - \mu' \sum_{i}(n_{i\uparrow} + n_{i\downarrow}) - \frac{h'}{2}\sum_{i}(n_{i\uparrow} - n_{i\downarrow}),
\end{eqnarray}
with 
\begin{eqnarray}
U'=-U, ~~\mu' = \frac{h-U}{2}, ~~h' = 2\mu -U,
\end{eqnarray}
where a constant term is dropped. Eq.(2) is transformed into
\begin{eqnarray}
%\rho(\mu',T) = 1 - 2\chi (T)(U' - 2\mu') + O[(U'-2\mu')^2]
& & 1 - \frac{const.}{T^{1/2}}\frac{1}{|\ln|U'-2\mu'||^{1/2}} \leq \rho(\mu',T) \nonumber \\ \hspace{0.5cm}
& & \leq 1 + \frac{const.}{T^{1/2}}\frac{1}{|\ln|U'-2\mu'||^{1/2}}
\end{eqnarray}
for small $|U'-2\mu'|$, where  $\rho = \frac{1}{M}\sum_i \langle n_{i\uparrow} + n_{i\downarrow} \rangle$, the
electron density per site. Inequality (6) is of basic importance. For the system described by the one-band Hubbard model (4)
on the square lattice, the density $\rho(\mu',T>0)$ is thus the continuous 
function of the chemical potential $\mu'$ near $\frac{U'}{2}$,
i.e., at small doping (note that at half-filling, $\mu' = \frac{U'}{2}$). 
This can be justified by noting the fact $\lim_{\delta \rightarrow 0} 
\rho(\mu'=U'/2 + \delta,T>0) = \lim_{\delta \rightarrow 0} 
\rho(\mu'=U'/2 - \delta,T>0)$. An alternative criterion for phase separation is based on
this fact\cite{daggo}: if a discontinuity is found in $\rho(\mu',T)$ as a 
function of $\mu'$, then the densities inside the gap are unstable, giving
rise to a phase-separated state, and if it is not found, then no phase separation
occurs. Actually, phase separation falls in the class of the first-order
phase transition,
as the
first-order phase transition into two phases with different densities is 
featured by
the discontinuity of $\rho(\mu',T)$. Thus the continuity of $\rho(\mu',T>0)$ near
$U'/2$
suggests that the one-band Hubbard model on the square lattice at $T>0$
does not exhibit phase separation at {\it small} doping (near half-filling), 
which is consistent 
with 
the quantum Monto Carlo calculation\cite{moreo,daggo} and with the result of 
high-temperature expansion for
the two dimensional t-J model with small values of $J/t$\cite{puti} which can be viewed as 
the strongly coupling limit
of the Hubbard model.  As in the beginning we 
do not fix the sign of $U$, this {\it exact} result remains true for both 
negative and 
positive $U'$ as well as 
the 
vanishing external field ($h'=0$). If the 
system is indeed paramagnetic, as 
being plausibly expected, then
this result could be extended to moderate dopings. One may observe that this 
approach only works for the two dimensions (or the one dimension) but not for the three
dimensions, as the Ghosh's results were
obtained only for the low dimensional cases.  

Now let us look at this problem when the system is doped away from half-filling.
We introduce the $\eta$ pairing (or pseudospin) operators
\cite{lieb,yang} as $\eta^{+} = \sum_{i} \epsilon(i) c_{i\uparrow}^{\dagger}c_{i\downarrow}^{\dagger}$, $\eta^{-} = (\eta^{+})^{\dagger}$, and $\eta^{z}=\frac12 (N - M)$ with $N = \sum_i (n_{i\uparrow}+n_{i\downarrow})$. They obey the usual $SU(2)$ Lie algebra. We from now
on set $h=0$ in Eq.(1). As a matter of fact, $[\eta^{-}, H] = 2 (\mu_c - \mu)\eta^{-}$, where
$\mu_c = \frac{U}{2}$. Using this commutator and the cyclicity under the trace
we obtain\cite{sz}
\begin{eqnarray}
F(\mu,T) \{ 1 - \exp [2 \beta (\mu_c - \mu)] \} = \rho (\mu, T) - 1,
\end{eqnarray}
with $F(\mu,T) = \frac{1}{M} \langle \eta^{+}\eta^{-} \rangle $, 
and $\beta$ the inverse temperature. Since the right-hand side of (7) is an 
intensive quantity, $F(\mu,T)$, being obviously non-negative, should also
be intensive. Furthermore, it is well-known\cite{shiba} that, 
based on the particle-hole transformation,
  $\rho (\mu, T) = 1$ (half-filling) as $\mu = \mu_c$ and vice versa.
Consequently, combining this result and Eq.(7) we get the following relation
\begin{eqnarray}
& & \rho (\mu, T) > 1 ~as~ \mu > \mu_c, ~~\rho (\mu, T) = 1 ~as~ \mu = \mu_c, \nonumber \\
& & and ~~\rho (\mu, T) < 1 ~as~ \mu < \mu_c.
\end{eqnarray} 
This {\it exact} constraint may have some implications. First, a direct 
consequence is $F(\mu, T)>0$ for $\mu \neq \mu_c$. Second, since $\rho (\mu, T)$ is analytic in the neighborhood
of $\mu = \mu_c$ (ensured by (6)), at small doping, we can
expand $\rho (\mu, T)$ in powers of $(\mu - \mu_c)$, obtaining
\begin{eqnarray}
 \rho (\mu, T) = 1 + \frac{\partial \rho (\mu, T)}{\partial \mu}|_{\mu = 
\mu_c} (\mu - \mu_c)  
+ O[(\mu - \mu_c)^2],  
\end{eqnarray} 
where we have used $\rho (\mu=\mu_c, T) = 1$. Up to the second order in 
$(\mu - \mu_c)$, we see that 
$\frac{\partial \rho (\mu, T)}{\partial \mu}|_{\mu = \mu_c}= finite > 0$ 
(otherwise it contradicts (6) and (8)). 

Since the functions $\rho (\mu, T)$ and $F(\mu, T)$ are closely related,
 let us now study the properties of the latter. 
From Eq.(7) we know that $F(\mu, T)$ is finite,
and comply
\begin{eqnarray}
0 <  F(\mu, T) \leq \frac{1}{|1-\exp [2 \beta (\mu_c - \mu)]|}
\end{eqnarray}
for $\mu \neq \mu_c$.
Furthermore, in accordance with the definition of $F(\mu, T)$, we have
\begin{eqnarray}
F(\mu, T) \equiv \frac{1}{M} \frac{Tr[e^{-\beta (H_0 -\mu N)}\eta^+\eta^-]}{Tre^{-\beta (H_0 -\mu N)}},
\end{eqnarray}
where we have denoted Eq.(1) by $H = H_0 -\mu N$ (Recall that $h=0$). Employing
the following symmetric particle-hole transformation,
\begin{eqnarray}
c_{i\uparrow}^{\dagger} \rightarrow \epsilon(i) c_{i\uparrow}, ~~~c_{i\downarrow}^{\dagger} \rightarrow \epsilon(i) c_{i\downarrow},
\end{eqnarray}
we find $H \rightarrow {\tilde H} = H_0 - (U-\mu)N + (U-2\mu)M$, and $\eta^+\eta^- \rightarrow \eta^-\eta^+$. By applying the unitary transformation, Eq.(12), to $F(\mu, T)$, we obtain
\begin{eqnarray} 
& & F(\mu, T) = \frac{1}{M} \frac{Tr\{e^{-\beta [H_0 -(U-\mu)N]}\eta^-\eta^+\}}{Tr e^{-\beta [H_0 -(U-\mu)N]}} \nonumber \\ 
& & = F(2\mu_c -\mu,T) e^{2\beta (\mu - \mu_c)},
\end{eqnarray}
where we have used $\eta^-\eta^+ = \eta^+\eta^- - 2 \eta^z$, the definition of
$F(\mu, T)$, as well as Eq.(7). Eq.(13) completely determines the form of $F(\mu, T)$. 
 As $\mu'= \mu_c -\mu$, it becomes $F(\mu_c -\mu,T)
= F(\mu_c + \mu,T) \exp (-2\beta \mu)$, which reflects the symmetry of $F(\mu, T)$ as
a function of $\mu$. Similarly, we can obtain, after 
operating the transformation (12) to $\rho(\mu,T)$, the following equation
\begin{eqnarray}
\rho(\mu,T) = 2 - \rho(2\mu_c - \mu,T).
\end{eqnarray}
One may verify that Eqs.(7), (13) and (14) are self-consistent.

By differentiating Eq.(11) with respect to $\mu$, and using the unitary 
particle-hole transformation (12) to the thermal averages involved, and then 
noting Eqs.(7) and (13), one can 
prove exactly the following expression
\begin{eqnarray}  
& & \frac{\partial F(\mu,T)}{\partial \mu} - \frac{\partial F(2\mu_c - \mu,T)}{\partial \mu} \nonumber \\
& & = \frac{\beta}{M} (\langle N^2 \rangle - \langle N \rangle^2)_{2\mu_c - \mu,T},
\end{eqnarray}
where $\langle \cdots \rangle_{\mu,T} $ means the thermal average with the chemical
potential $\mu$ at temperature $T$. Considering $\langle N^2 \rangle - \langle N \rangle^2 \geq 0$, we get $\frac{\partial F(\mu,T)}{\partial \mu} \geq \frac{\partial F(2\mu_c - \mu,T)}{\partial \mu}$. By (14) and (10), we obtain the
inequality
\begin{eqnarray}  
\frac{\partial F(\mu,T)}{\partial \mu} \leq \frac{2\beta e^{2\beta(\mu_c-\mu)}}{(e^{2\beta(\mu_c-\mu)}-1)^2}
\end{eqnarray}
for $\mu < \mu_c$. 

Till now we have already had some basic knowledge about the function $F(\mu,T)$, {\it i.e.}, it should satisfy (7), (10), (13), (15) and (16) simultaneously. 
Under these conditions, we can solve equation (13) exactly, and the solution, being surprisingly
simple, is given by
\begin{eqnarray}
F(\mu,T) = \frac{C(T)}{1 + e^{2\beta(\mu_c-\mu)}},
\end{eqnarray}
where $C(T)$ is a positive, finite constant at temperature $T$, satisfying
\begin{eqnarray}
0 < C(T) \leq 1.
\end{eqnarray}
We like to mention here that we can not obtain the 
closed form of $C(T)$ for the time being, but such a form is enough for our 
purpose. Consequently, by Eq.(7), we have
\begin{eqnarray}
\rho(\mu,T) = 1 + C(T) \tanh [\beta (\mu -\mu_c)].
\end{eqnarray}
One may check that the solutions given by Eqs.(17) and (19) satisfy all 
aforementioned properties of functions $F(\mu,T)$ and $\rho(\mu,T)$. 

With these facts, we are ready to address the problem of phase separation away
from half-filling. As we have obtained an exact, 
explicit expression for $\rho(\mu,T)$, say, Eq.(19), it is obvious that $\rho(\mu,T)$ as a
function of $\mu$ is continuous at any fillings.  
This result shows that there is no occurrence of the first-order phase 
transition, which in turn leads to the
conclusion that the one-band Hubbard model can not exhibit the phenomenon of 
phase separation on a square lattice at finite temperature, consistent
 with the numerical calculations\cite{moreo,daggo}. 
 
A few remarks concerning Eqs.(17) and (19) are in order. (1) As one may note
that, Eq.(19) looks to be reasonable, as it comes directly from the symmetry
of the Hubbard model, and is well in agreement with the inequality (6). 
Moreover, $\rho(\mu,T)$, given by (19), has the qualitatively similar behaviors
compared with numerical results on square lattices with small sizes, though a
quantitative comparison is impossible due to uncertainty of $C(T)$ as well as
the lack of accurate data. Clearly, Eq.(19) remains valid in the thermodynamic
limit. We like to mention here that we can not prove the uniqueness of the
solution Eq.(17), but the other solutions, if exist, might have the similar
forms as Eq.(17)\cite{schad}, which would not affect our conclusion. (2) As indicated in Eq.(19), $\rho(\mu,T)$ might have different 
behaviors for positive and negative $U$ because of possible $\mu_c-$dependence of 
$C(T)$. 
(3) We emphasize once again that Eqs.(17) and (19) work only for finite
temperatures and finite chemical potentials. One can not extract any useful
information for the zero-temperature case from this study. Based upon the 
present result, however, we
could say that it may be inappropriate to use the one-band Hubbard model as a
model to explain the phenomena of phase separation observed in high Tc
superconductors, at least at finite temperatures.

In summary, we show in a somewhat exact way that, using the known exact results and 
symmetry properties for the one-band Hubbard
model, there is 
no phase separation at any fillings on a square 
lattice at finite temperature for both negative and positive U. 
This result is consistent with 
the quantum Monto Carlo calculation and some analytic result, and is also compatible with the recent 
exact result obtained in Ref.\cite{azn}. Nevertheless, we like to mention that
our result does not cover the case at zero temperature. As there are also 
strong numerical evidences showing no occurrence of phase separation at zero temperature, 
how to obtain an exact proof for both $U>0$ and $U<0$ is still a fascinating topic. 

\acknowledgements

The author is indebted to Dr. A. Schadschneider and Prof. B.H. Zhao for various
discussions. He is also grateful to Prof. J. Zittartz and ITP of Universit\"at zu 
K\"oln for the warm hospitality, and to the Alexander von Humboldt Stiftung for financial
support.

\end{document}